\documentclass{PoS}
\usepackage{wrapfig}
\usepackage{amssymb,amsmath,array}

\title{Transverse $\Lambda$ polarization at high energy colliders }

\ShortTitle{Transverse $\Lambda$ polarization at high energy colliders}

\author{\speaker{Dani\"{e}l Boer}\\ %\thanks{A footnote may follow.}\\
        Theory Group, KVI, University of Groningen\\
        Zernikelaan 25, NL-9747 AA Groningen, The Netherlands\\
        E-mail: \email{D.Boer@rug.nl}}
      
\abstract{Measurements of transverse polarization of $\Lambda$ hyperons 
produced in high energy $p \, p$ collisions may help to address several open 
issues about $\Lambda$ production and polarization mechanisms,  
such as the amount of $SU(3)$ breaking, the importance of gluons and
sea quarks, and the origin of spontaneous $\Lambda$ polarization. 
The process $p + p \to \Lambda^\uparrow +
  \text{jet} + X$ at midrapidity is ideally suited for this purpose, for
  instance at LHC's ALICE experiment. 
New expressions and predictions are presented for 
the transverse $\Lambda$ polarization in this process, 
within a factorized description 
which involves transverse momentum and spin dependence in the
fragmentation process. Uncertainties from the unpolarized
  $\Lambda$ fragmentation functions, due to the unknown magnitude of $SU(3)$ 
breaking and the apparent inconsistency between $pp$ and $e^+ e^-$ data, are
  investigated. }

\FullConference{XVIII International Workshop on Deep-Inelastic Scattering and Related Subjects, DIS 2010\\
                April 19-23, 2010\\
                Firenze, Italy}

\begin{document}

\section{Introduction}
It is well-known since the mid-1970's that $\Lambda$ hyperons produced
in unpolarized $p\, p$ collisions are to a large degree polarized
transversely to the production plane \cite{Lesnik:1975my}.  There have
been many experimental and theoretical investigations aimed at
understanding this striking polarization phenomenon, but no consensus
has been reached about its origin. One of the difficulties in
interpreting the available (mostly fixed target) data is that they are
not or only partly in a region where a factorized description of the
cross section is expected to be applicable. For a comprehensive review
of these relatively low energy data ($\sqrt{s} \leq 62$ GeV) see Ref.\ 
\cite{Panagiotou:1989sv}. New high-energy hadron collider data would
be very welcome, for instance from RHIC, Tevatron, or LHC, but there
the capabilities to measure the $\Lambda$ polarization $P_\Lambda$ via
the self-analyzing parity violating decay $\Lambda \to p \, \pi^-$ are
typically restricted to the midrapidity region ($\eta \approx 0$), 
where protons can be 
identified\footnote{An alternative may be to use neutral decays 
$\Lambda \to n\, \pi^0$ (50\% less frequent than $p
\pi^-$) \cite{Cork:1960zz}.}. For symmetry reasons $P_\Lambda=0$ at
$\eta=0$ in $p\, p$ collisions in the center of mass frame, hence 
the degree of transverse polarization $P_\Lambda$ around midrapidity
is expected to be very small. As an
alternative to $p + p \rightarrow \Lambda^{\uparrow} + X$, it has been
suggested \cite{Boer:2007nh} to perform $\Lambda$ polarization studies
in the process $p + p \to (\Lambda^\uparrow \text{jet}) + \text{jet} +
X$, where the $\Lambda$ and jets can be measured in the midrapidity region
without paying a suppression penalty. It is especially of interest at
LHC, where the factorized description is expected to apply and certain
simplifications may arise due to the large center of mass energy
$\sqrt{s}$. 

\section{Spontaneous $\Lambda$ polarization in a factorized approach}
The magnitude of the $\Lambda$ polarization in $p + p \rightarrow
\Lambda^{\uparrow} + X$ at large $x_F$ remains large up to the highest measured
transverse momentum $p_T$ of the $\Lambda$: $p_T \sim 4$ GeV/$c$. For
sufficiently high $p_T^{}$, perturbative QCD and collinear
factorization should become applicable. 
Consider for example the $q g \to q g$ subprocess contribution to
the $p + p \to \Lambda + X$ cross section in collinear
factorization. It is of the form:
$\sigma \sim {q(x_1)} \otimes {g(x_2)}  \otimes
\hat{\sigma}_{q g \to q g} \otimes {D_{\Lambda/q}(z)}$, where
${q(x_1)}$ is the quark density in proton 1, ${g(x_2)}$ is the gluon
density in proton 2, and 
${D_{\Lambda/q}(z)}$ is the $q \to \Lambda$ fragmentation function
(FF). In a
similar way the transverse polarization should be of the form:
$P_{\Lambda} \sim q(x_1)  \otimes g(x_2) \otimes
\hat{\sigma}_{q g \to q g} \otimes {?}$, involving the unpolarized
parton densities and the {\it unpolarized} hard partonic subprocess 
because $\Lambda$ polarization created
in the hard partonic scattering is very small, 
$P_\Lambda \sim \alpha_s m_q/\sqrt{\hat s}$ \cite{Kane:1978nd}.
The question mark indicates
that for symmetry reasons at leading twist 
there is no collinear fragmentation function
describing $q \to \Lambda^\uparrow X$. In
collinear factorization $P_{\Lambda}$ is thus necessarily power
suppressed. 
Dropping the demand of {\em collinear\/} factorization, 
does allow for a leading twist solution: the transverse momentum
dependent (TMD) fragmentation function 
$D_{1T}^\perp(z,{\mathbf{k}_T})$ \cite{Mulders:1995dh} for
$\Lambda$'s. It describes a nonperturbative 
{$\mathbf{k}_T \times \mathbf{S}_T$} dependence in the fragmentation
process, which is allowed by the symmetries (parity and time
reversal). As the $\Lambda$ polarization arises in the fragmentation of 
an {{\em unpolarized}} {quark}, it carries the descriptive name 
``polarizing fragmentation function'' \cite{Anselmino:2000vs}. 
${D_{1T}^\perp}$ has been extracted \cite{Anselmino:2000vs} 
from the low energy 
$p + p/Be \to \Lambda^{\uparrow}/\bar\Lambda^{\uparrow} + X$ data, 
where gluon FFs ($g \to \Lambda \, X$) are expected to be hardly
relevant. Reasonable valence quark FFs are obtained: 
$D_{1T}^\perp$ has opposite signs for $u/d$ versus 
$s$ quarks, and the latter is larger. This is responsible for the 
cancellations that ensure $P_{\; \bar \Lambda} \approx 0$.
The extraction has been done under the restriction of $p_{T} > 1$
  GeV/$c$, in order to exclude the soft regime but to retain
sufficient data to make a fit to. Whether this restriction 
is sufficiently strict is a matter of concern, due to the large 
$K$ factors required to obtain a cross section description. 
Data at higher $\sqrt{s}$ and $p_T$ would be much safer to ensure the validity
of the factorized description. As pointed out, these
do not necessarily require large $x_F$, if one goes beyond 
$p + p \rightarrow \Lambda^{\uparrow} + X$. 
If the origin of the transverse $\Lambda$ polarization
is indeed due to polarizing fragmentation, then another, related asymmetry
could be observed that does not need to vanish at $\eta_\Lambda=0$, 
namely in the process $p + p \to (\Lambda^\uparrow \text{jet}) +
  \text{jet} + X$ \cite{Boer:2007nh} and actually also in  
$p + p \to (\Lambda^\uparrow \text{jet}) + X$.

\section{Jet-$\Lambda^\uparrow$~production}

The suggestion of \cite{Boer:2007nh} 
is to select two-jet events and to measure
the jet momenta $K_j^{}$ and $K_{j'}^{}$ (with $K_{j}^{}\cdot
K_{j'}^{} = {\cal O}(\hat{s})$), in addition to the momentum
$K_\Lambda^{}$ and polarization $S_\Lambda^{}$ of the $\Lambda$ that
is part of either of the two jets. A single spin asymmetry proportional to
$\epsilon_{\mu\nu\alpha\beta} K_j^{\mu} K_{j'}^{\nu}
K_\Lambda^{\alpha} S_\Lambda^\beta$ can then arise, which is neither
power suppressed, nor needs to be zero at midrapidity. 
In the center of mass (c.o.m.) 
frame of the two jets the asymmetry is of the form:
\begin{equation}
\text{SSA}
=\frac{d\sigma({+}\mathbf{S}_\Lambda)\,
{-}\,d\sigma({-}\mathbf{S}_\Lambda)}
{d\sigma({+}\mathbf{S}_\Lambda)\,
{+}\,d\sigma({-}\mathbf{S}_\Lambda)}
=\frac{\hat{\mathbf{K}}{}_j{\cdot}
({\mathbf{K}_\Lambda{\times}\mathbf{S}_\Lambda})}{
z\,M_\Lambda}\,{\frac{d\sigma_T}{d\sigma_U}}\ 
\label{SSA}
\end{equation}
The analyzing power {$d\sigma_T/d\sigma_U$} of the asymmetry 
depends on $D_{1T}^\perp$. This new
$\Lambda$+jets observable could allow for a more trustworthy extraction
of $D_{1T}^\perp$ (for both quarks and gluons) and subsequent
predictions, for instance for
semi-inclusive DIS \cite{Anselmino:2001js}.

At RHIC and LHC this process $p \, p \to \left(
\Lambda^\uparrow \text{jet}\right) \, \text{jet} \, X$ can be studied.
For instance, the ALICE experiment has 
excellent PID capabilities which allow measurements of 
$\Lambda$'s over a wide $p_T$ range.
The ALICE rapidity coverage is ${-}0.9\,{\leq}\,\eta\,{\leq}\,{+}0.9$.
If the jet rapidities ($\eta_{j,j'}$) are in this range 
and if gluon fragmentation is at least
as important as quark fragmentation for both unpolarized and polarized
$\Lambda$ production, then the process
is dominated by gluon-gluon ($gg{\rightarrow}gg$) scattering\footnote{Unlike in Ref.\ \cite{Boer:2007nh}, here it will be assumed 
that universality of $D_{1T}^\perp$ holds throughout \cite{Metz:2002iz}.},
i.e.\ $d\sigma_T/d\sigma_U \approx
D_{1T}^{\perp\,g}(z{,}K_{\Lambda\,T}^2)/D_1^g(z{,}K_{\Lambda\,T}^2)$.
Because no model or fit for $D_{1T}^{\perp\,g}$ is available yet, no
predictions can be made in this case. 
If it happens that {$D_{1T}^{\perp\, g}\,{\ll}\,D_{1T}^{\perp\, q}$}, then
one can use the extracted $D_{1T}^{\perp\,q}$ to obtain an
estimate. Taking into account the $qg\to qg$ subprocess, one finds 
for $\eta_{j'}{\approx}\,{-}\eta_j \approx 0$ 
\begin{equation}
\frac{d\sigma_T}{d\sigma_U}
\approx \frac{\sum_q f_1^q(x)D_{1T}^{\perp\,q}(z{,}K_{\Lambda\,T}^2)}
{\sum_q f_1^q(x)(D_1^q(z{,}K_{\Lambda\,T}^2)
+D_1^g(z{,}K_{\Lambda\,T}^2))+f_1^g(x) D_1^g(z{,}K_{\Lambda\,T}^2)/0.4},
\label{SigTSigU}
\end{equation}
where $x\,{\approx}\,2|\mathbf{K}_{j\perp}|/\sqrt{s}$. The factor
$0.4$ comes from the ratio $d\hat\sigma_{qg\rightarrow
  qg}/d\hat\sigma_{gg\rightarrow gg}$ at midrapidity; other partonic channels 
can be neglected \cite{Boer:2007nh}.
\begin{figure}[htb]
\includegraphics[width=0.5\textwidth]{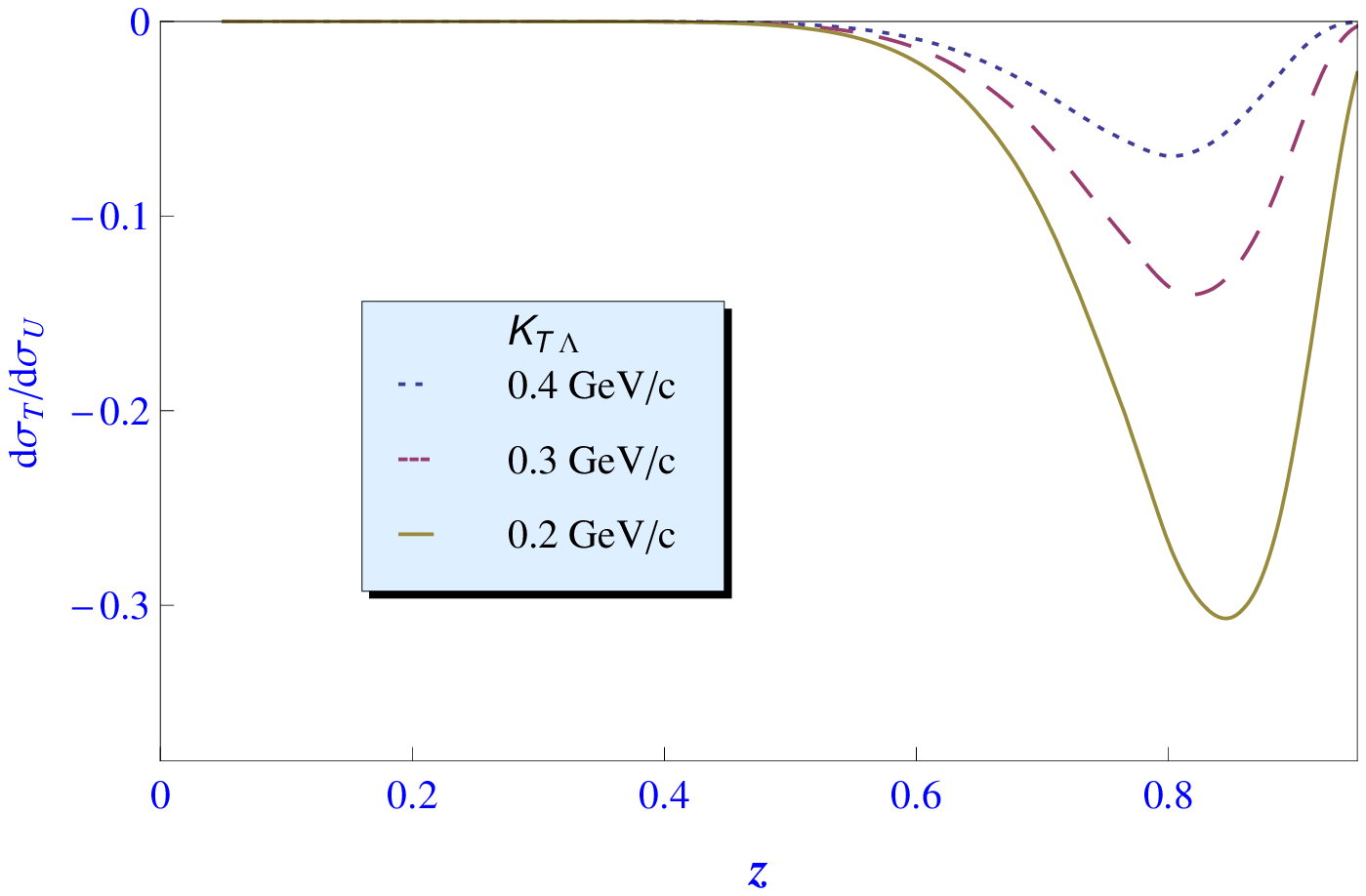}
\includegraphics[width=0.5\textwidth]{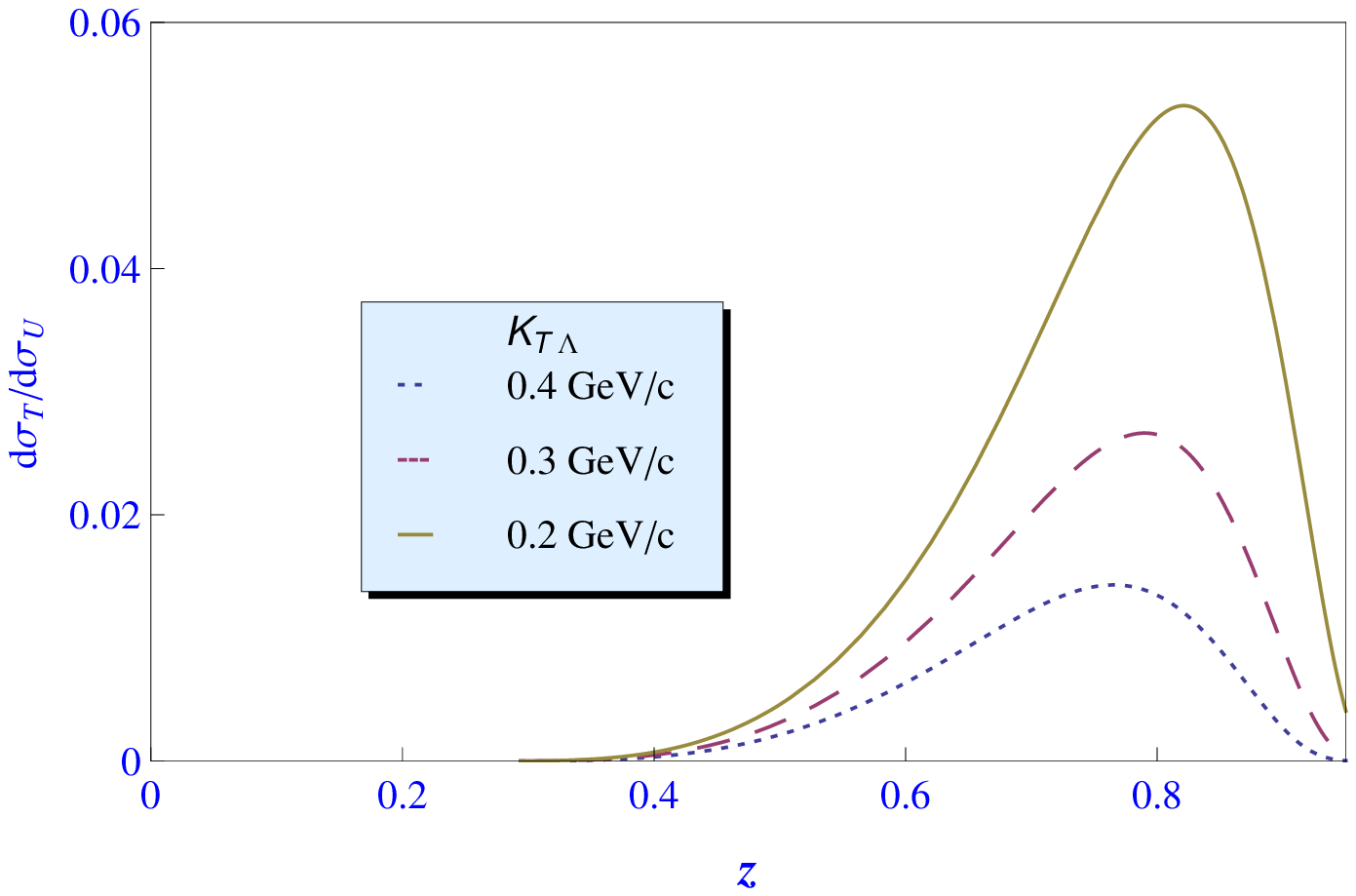}
\caption{The asymmetry $d\sigma_T/d\sigma_U$ for $\eta_{j},\eta_{j'} =
  0$, $|K_{j\, \perp}|=|K_{j'\, \perp}| = 70 $ GeV, $\sqrt{s}=14$ TeV,
  using DSV (left) and IMR (right) FFs.}
\label{Asymm}
\end{figure}
For estimates we will use the DSV \cite{deFlorian:1997zj}
and IMR \cite{Indumathi:1998am} unpolarized FFs sets, and their accompanying
$D_{1T}^\perp$ parametrizations from Ref.\ \cite{Anselmino:2001js}. 
Both DSV and IMR satisfy $D_1^g \ll D_1^q$ at larger $z$, as both
have been obtained from fits to $e^+ e^- \to \Lambda \, X$ data only, 
which is not sensitive to the fragmentation process $g \to \Lambda \, X$. 
The resulting asymmetry (\ref{SigTSigU}) evaluations 
are given in Fig.\ \ref{Asymm} for three different values of the
$\Lambda$ momentum component transverse to the jet direction (see
\cite{dis2009} for additional comments).
Very small asymmetries are obtained at smaller $z$, due to 
$D_{1T}^\perp$ being fitted to low energy
data where mostly valence quarks matter. 
This behavior need not be realistic at high energies. 
ALICE would have most data in the region $z <0.5$ and could therefore 
provide valuable information on sea quark and gluonic contributions.  

The asymmetry is quite sensitive to the
cancellation between $u/d$ and $s$ contributions, like in SIDIS
\cite{Anselmino:2001js}, and can even flip its overall sign depending 
on the amount of $SU(3)$ breaking in the unpolarized fragmentation 
functions: IMR includes $SU(3)$ breaking, whereas DSV does not. 
This aspect represents a large uncertainty, but more reliable estimates are 
not possible at this stage. ALICE may help to determine the importance 
of $SU(3)$ breaking and flavor cancellations. 

\begin{figure}[htb]
\includegraphics[width=0.32\textwidth]{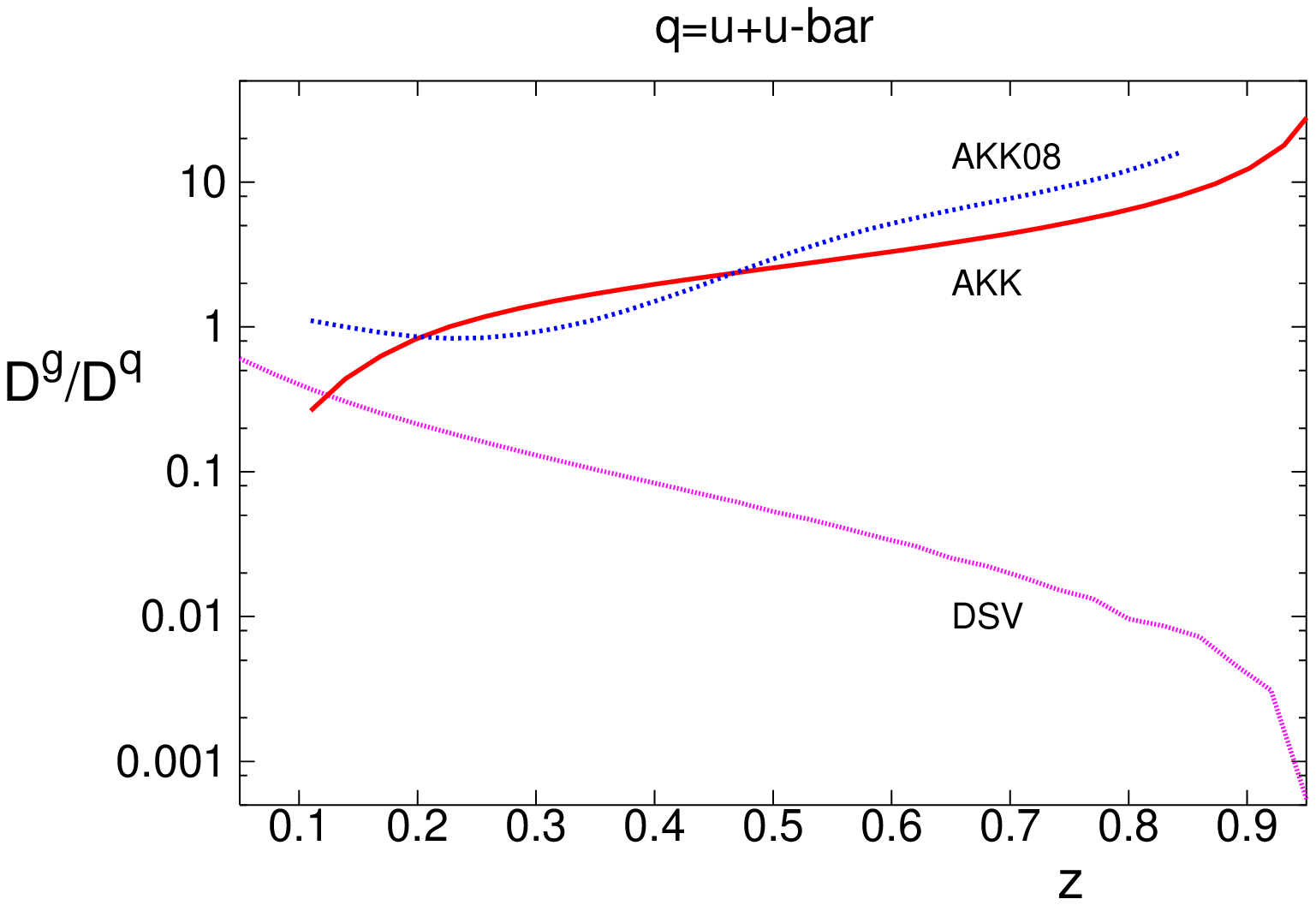}
\includegraphics[width=0.32\textwidth]{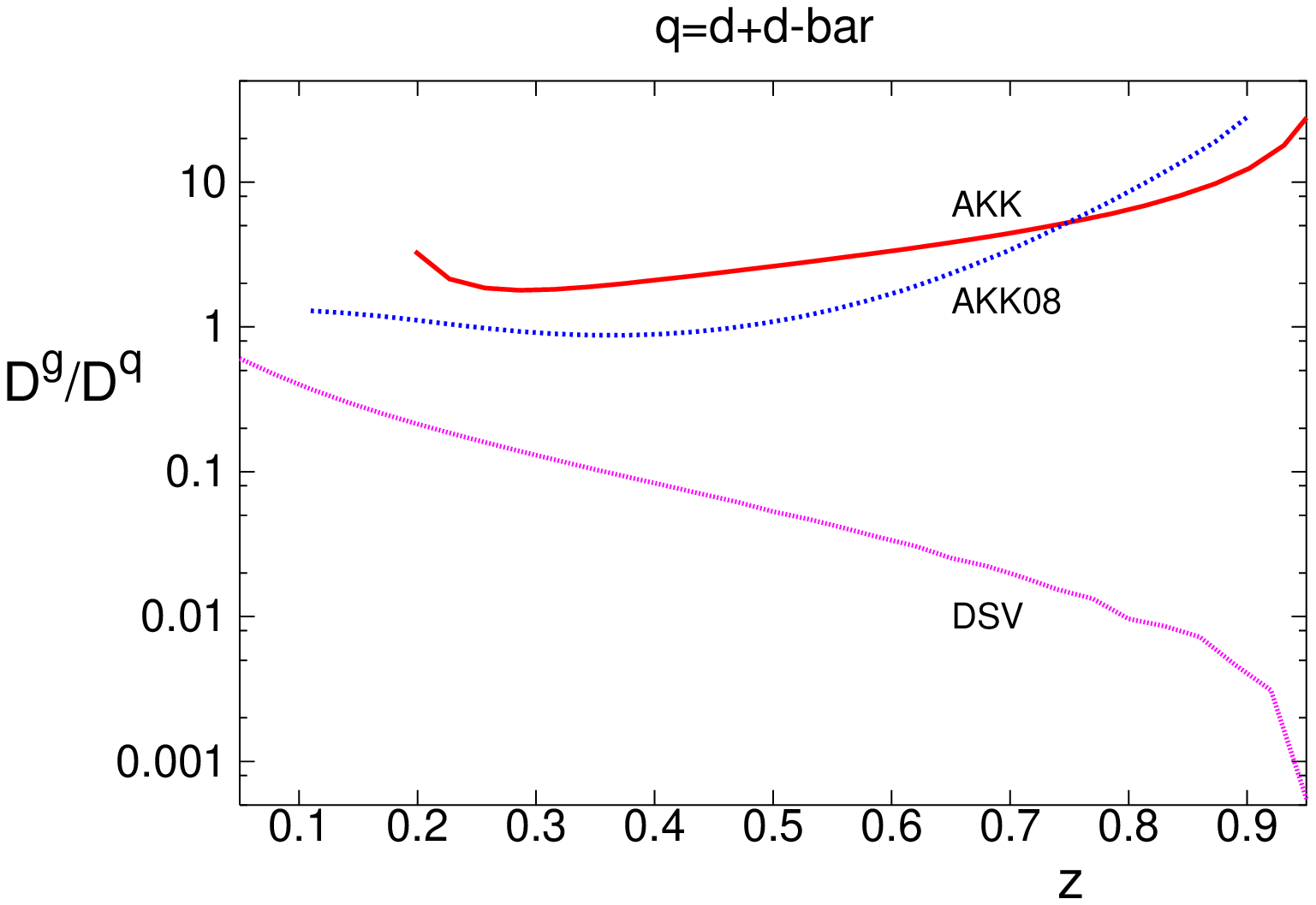}
\includegraphics[width=0.32\textwidth]{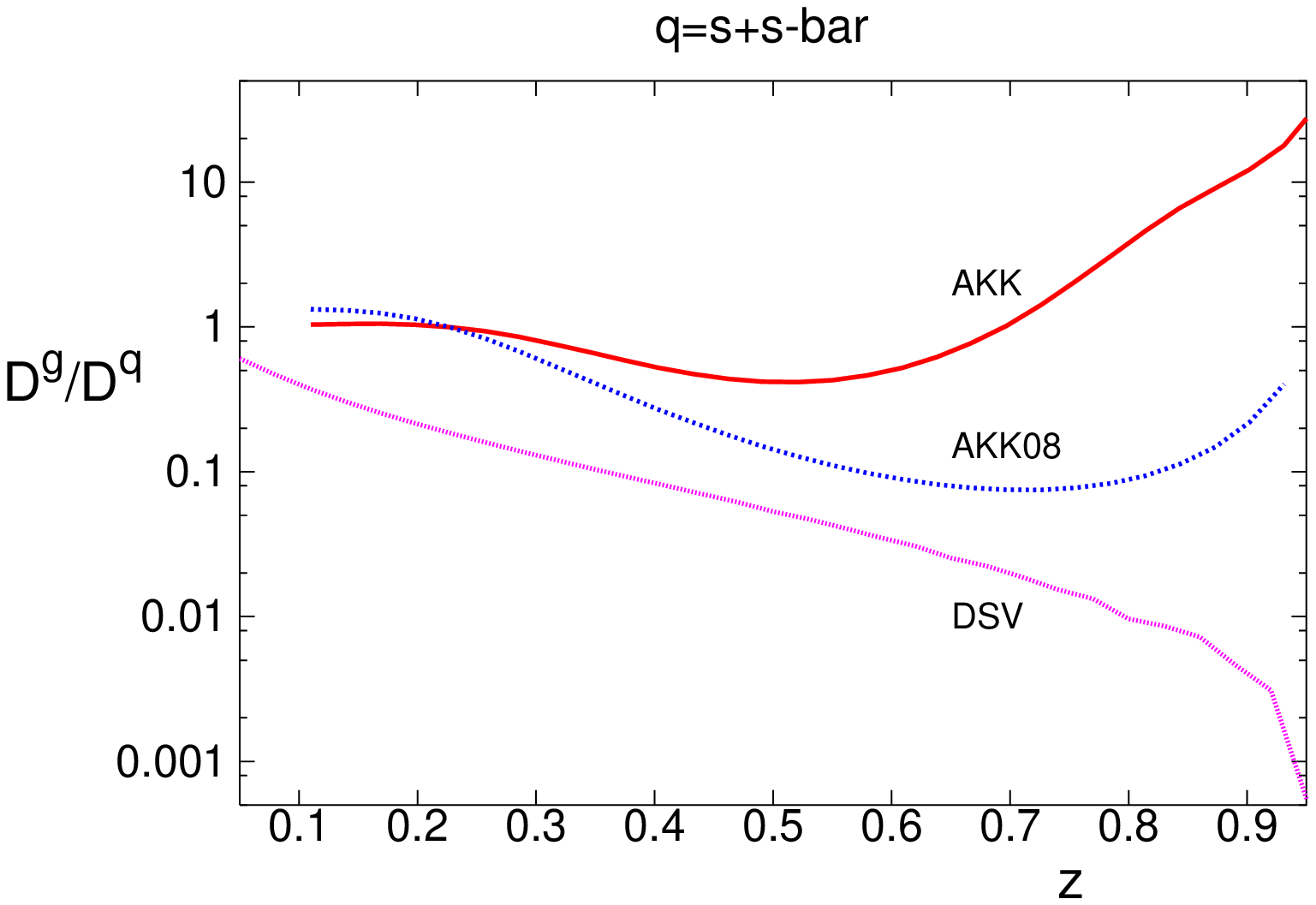}
\caption{The ratio $D_1^g/D_1^{q+\bar{q}}$ as function of $z$ 
for $q=u,d,s,$ respectively, for the leading order DSV, AKK and AKK08 sets, 
at the scale $Q=10$ GeV.} 
\label{Ratios}
\end{figure}

Unpolarized $\Lambda$ FFs have also been obtained 
taking into account midrapidity 
hadronic $\Lambda$ production data from STAR at RHIC: the AKK
\cite{Albino:2005mv} and AKK08 \cite{Albino:2007ns} sets. These have
very different characteristics compared to fits to $e^+ e^-$ data only. 
Fig.~\ref{Ratios} shows the ratio $D_1^g/D_1^{q+\bar{q}}$ for
$q=u,d,s$, respectively, for DSV, AKK and AKK08. The curves
differ widely and represent yet another uncertainty in the
predictions of (\ref{SigTSigU}). 
No $D_{1T}^\perp$ extraction has been performed with AKK or AKK08. 
A problem with that is that the most recent fit AKK08 considerably
undershoots the cross section of 
$p \, p \to \Lambda/\overline{\Lambda} \, X$ at midrapidity, even at 
$\sqrt{s}=200$ GeV and transverse momenta in the range 
$2 \leq \mathbf{p}_T \leq 5$ GeV/$c$. In
\cite{Albino:2007ns} it was therefore concluded that there is
``a possible inconsistency between the $p\, p$ and $e^+e^-$ reaction
data for $\Lambda/\overline\Lambda$ production'' (cf.\ Fig.~5 of
\cite{Albino:2007ns}). This issue ought to be clarified first.

The suggestion of \cite{Boer:2007nh} requires the opposite side jet
$j'$ to be measured, such that the jet-jet or equivalently the
partonic c.o.m.\ frame can be selected. 
However, if universality of $D_{1T}^\perp$ and gauge
invariance of the factorized expression are taken to hold, 
then one can actually consider 
$p \, p \to \left(\Lambda^\uparrow \text{jet}\right) \, X$,
increasing the statistics considerably. The asymmetry 
is also given as in Eq.\ (\ref{SSA}), but now in the
{\it hadronic} c.o.m.\ frame, which
can be shown by considering the Sudakov decomposition of the fragmenting
quark momentum $k=K_\Lambda/z+\sigma_\Lambda n_\Lambda+k_T$ 
(but not of the other 
momenta), with the uncommon choice $n_\Lambda=P_1+P_2$. Due to the large
$\sqrt{s}$ at the LHC, $\sigma_\Lambda$ is still suitably 
suppressed (at least by a factor $E_j/\sqrt{s}$), 
despite $n_\Lambda$ being non-lightlike.  
The asymmetry is now $\propto \epsilon_{\mu\nu\alpha\beta} K_j^{\mu}
(P_1+P_2)^{\nu} K_\Lambda^{\alpha} S_\Lambda^\beta$, which in the
hadronic c.o.m.\ frame indeed becomes $\propto \mathbf{K}_j{\cdot}
(\mathbf{K}_\Lambda{\times}\mathbf{S}_\Lambda)$. Eq.\ (\ref{SigTSigU})
is now replaced by 
\begin{equation}
\frac{d\sigma_T}{d\sigma_U} 
\approx 
\frac{
\int \frac{dy}{y}\sum_q \left(f^{qg}+f^{gq}\right) 
d\hat{\sigma}_{qg} D_{1T}^{\perp\,q}}{\int \frac{dy}{y}  
\left[ \sum_q \left(f^{qg}+f^{gq}\right)
d\hat{\sigma}_{qg} \left(D_1^q + D_1^g\right) 
+ f^{gg}d\hat{\sigma}_{gg} D_1^g \right]} , 
\label{SigTSigUnew}
\end{equation}
where the arguments of the fragmentation functions have been
suppressed and we introduced the notation: $f^{ab}\equiv x_1 f_1^a(x_1) 
x_2 f_1^b(x_2)$, $d\hat\sigma_{ab}\equiv d\hat\sigma_{ab\to ab}(y)+
d\hat\sigma_{ab\to ab}(1-y)$ with $y=-\hat t /\hat s$. 
The momentum fractions are fixed to be 
$x_1=x_\perp e^\eta/(2(1-y))$ and $x_2=x_\perp e^{-\eta}/(2y)$ for
given observed $x_\perp=2|\mathbf{K}_{j\perp}|/\sqrt{s}$ and $\eta=\eta_{j}$.
The $y$ integration is between 
$y_{\min}=x_\perp e^{-\eta}/2$ and $y_{\max}=1- x_\perp e^\eta/2$. The
expressions for the partonic cross sections can be found in Ref.\
\cite{Boer:2007nh}. For $D_1^g \gg D_1^q$ the last term in the denominator
is the dominant one: for $x_\perp=0.01, \eta=0$ it is
approximately a factor of 3 larger than the other term proportional to $D_1^g$.

\section*{Acknowledgments}

I thank Markus Diehl, Dae Sung Hwang, Piet Mulders, Pasquale Di Nezza,
Werner Vogelsang, and Feng Yuan for useful discussions.

\end{document}